\documentclass[twocolumn,prb]{revtex4-1}
\usepackage{graphicx,color}

\begin{document}

\title{First-principles calculations on the spin anomalous Hall effect of ferromagnetic alloys}

\author{Yoshio Miura$^{1,2}$}
\email{miura.yoshio@nims.go.jp}

\author{Keisuke Masuda$^{1}$}

\affiliation{$^{1}$Research Center for Magnetic and Spintronic Materials(CMSM), National Institute for Materials Science(NIMS), 1-2-1 Sengen,Tsukuba, Ibaraki, 305-0047, Japan}
\affiliation{$^{2}$Center for Spintronics Research Network (CSRN), Graduate School of Engineering Science, Osaka University,
Machikaneyama 1-3, Toyonaka, Osaka 560-8531, Japan}

\date{\today}

\begin{abstract}
The spin anomalous Hall effect (SAHE) in ferromagnetic metals, which can generate spin-orbit torque to rotate the magnetization of another ferromagnetic layer through a non-magnetic spacer in magnetic junctions, has attracted much attention.
We theoretically investigated the spin anomalous Hall conductivity (SAHC) of the L1$_0$-type alloys $X$Pt($X$=Fe,Co,Ni) on the basis of first-principles density functional theory and linear response theory.
We found that the SAHC of FePt is much smaller than the anomalous Hall conductivity (AHC), leading to very small polarization for the anomalous Hall effect $\zeta$=SAHC/AHC of around 0.1.  
On the other hand, the SAHC increases with increasing number of valence electrons($N_{\rm v}$), and  CoPt and NiPt show relatively large values of $|\zeta|$, greater than 1.
The negative contribution of the spin-down-down component of AHC is the origin of the large SAHC and $\zeta$ in CoPt and NiPt, which is due to the anti-bonding states of Pt around the Fermi level in the minority-spin states.

\end{abstract}

\maketitle

Reducing the power consumption for magnetization reversal in magnetic tunnel junctions (MTJs) is one of the most important tasks for the realization of magnetic random access memory (MRAM)\cite{2017Bhatti-MaterialsToday}. Current-induced switching, such as spin transfer torque (STT)\cite{1989Slonczewski-PRB,1996Slonczewski-JMMM,1996Berger-PRB,1999Myers-Science}-induced switching, has emerged as a promising method for magnetization reversal in MTJs. In STT-induced switching, the spin flip of a conduction electron flowing perpendicular to the plane gives a torque to the local spin moment due to the conservation of angular momentum, and continuous spin-flip scattering of conduction electrons can rotate the magnetization direction of a free layer in MTJs. In order to realize ultra-high density MRAM, we must reduce the critical current density $J_{\rm c_0}$ in the magnetization reversal by STT up to 10$^5$ A/cm$^2$\cite{2017Bhatti-MaterialsToday}. 

Spin-orbit torque (SOT)\cite{2011Miron-Nature,2012Liu-Science,2016Fukami-NatureNano}-induced switching also has attracted much attention in recent years because it enables high-speed and reliable operation in three-terminal MRAM, where the magnetization of a free layer can be switched by an in-plane current flowing in the non-magnetic layer attached to the free layer in the current-in-plane geometry.   SOT is caused by spin current injection into a ferromagnetic layer from a non-magnetic layer due to the spin-Hall effect (SHE) in nonmagnetic metals\cite{1999Hirsch-PRL, 2013Hoffman-IEEETransMagn}.  The spin current $\mbox{\boldmath $J$}_{\rm s}$ generated by the SHE can be given by $\mbox{\boldmath $J$}_{\rm s} \propto \alpha_{\rm SH}[ \mbox{\boldmath $\hat{s}$} \times \mbox{\boldmath $J$}_{\rm c} ]$, where $\alpha_{\rm SH}$ is the spin Hall angle given by the ratio of the spin Hall conductivity $\sigma ^{\rm spin}_{\rm xy}$ to the conductivity of the charge current $\sigma _{\rm xx}$, i.e., $\sigma ^{\rm spin}_{\rm xy}/\sigma _{\rm xx}$, \mbox{\boldmath $\hat{s}$} is the direction of the quantization axis of an electron spin, and $\mbox{\boldmath $J$}_{\rm c}$ is the charge current. To obtain efficient spin-orbit torque in ferromagnetic layers, a large spin-orbit interaction is necessary in the non-magnetic layer. Thus, heavy metals such as Pt and Ta have been used as non-magnetic under-layers. While SOT-induced switching leads to more efficient magnetization switching in MRAM, the direction of the spin quantum axis of injected electrons related to the direction of the torque is limited by the geometry of the device, such as the direction of the current flow and the magnetic anisotropy of the ferromagnetic layers. 

Recently, Taniguchi {\it et al.} proposed a new type of SOT-induced switching\cite{2015Taniguchi-PRAp}, in which the non-magnetic layer is replaced by a ferromagnetic layer, and spin current due to the spin anomalous Hall effect (SAHE) in the ferromagnetic layer can provide torque to rotate magnetization of the ferromagnetic upper layer through the non-magnetic spacer. Figures 1(a) and (b) show schematic viewgraphs of SAHE and SHE in ferromagnetic materials (see Supplemental Materials for a detailed explanation on the definition of SAHE\cite{SM1}).  In this case, the spin current $\mbox{\boldmath $J$}_{\rm s}$ generated by the SAHE can be given by $\mbox{\boldmath $J$}_{\rm s} \propto (\zeta-\beta)\alpha_{\rm AH} [ \mbox{\boldmath $\hat{m}$} \times \mbox{\boldmath $J$}_{\rm c}]$, where \mbox{\boldmath $\hat{m}$} is the direction of magnetization of the bottom ferromagnetic layer, $\alpha_{\rm AH}$ is the anomalous Hall angle $\sigma _{\rm xy}/\sigma _{\rm xx}$ , $\beta$ is the spin-polarization of the bottom ferromagnetic layer, and $\zeta$ is the ratio of the spin current to the transverse charge current by the AHE, i.e., $\sigma ^{\rm spin}_{\rm xy}/\sigma _{\rm xy}$. Thus, ferromagnetic materials showing large $\zeta$ are promising as high-efficiency spin current sources by the SAHE .  So far, several experiments have been carried out to evaluate the efficiency of the spin-orbit torque originating from the spin current in ferromagnetic metals for various systems\cite{2017PRB-Das,2018Iihama-NatureElec,2018Gibbons-PRAp,2018Bose-PRAp,2019Seki-PRB}. 
Theoretical studies on the SAHE of typical ferromagnetic metals such as Fe, Ni and Co have been performed using first-principles calculations\cite{2019Amin-PRB}. Furthermore, the correlation between AHE and SHE of CoPt was theoretically investigated\cite{2021Qu-JPSJ}. However, the material dependence of SAHE in ferromagnetic alloys is still unclear. 
Here, we investigate and discuss the efficiency of SAHE on the basis of first-principles calculations and linear response theory in order to clarify the possible origin of the SAHE. To this end, we focus on the Pt-based L1$_0$ binary alloys, FePt, CoPt and NiPt, as sources of SOT in ferromagnetic materials.


\begin{figure}[tp]
\includegraphics[height=0.18\textheight,width=0.49\textwidth]{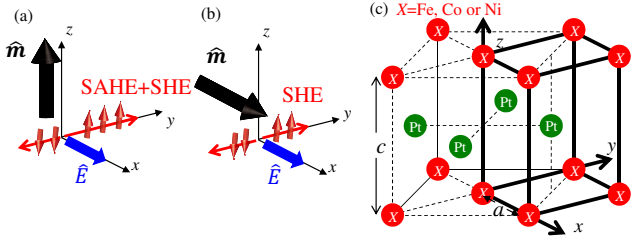}
\caption{Schematic viewgraph of (a) spin anomalous Hall effect (SAHE: magnetization \mbox{\boldmath $\hat{m}$} is perpendicular to electric field \mbox{\boldmath $\hat{E}$}) and (b) magnetization independent spin Hall effect (SHE: \mbox{\boldmath $\hat{m}$} is parallel to \mbox{\boldmath $\hat{E}$}) in ferromagnets. (c)Crystal structure and coordinate system of L1$_0$-$X$Pt($X$=Fe,Co,Ni). 
}
\end{figure}

We performed density functional theory (DFT) calculations on L1$_0$-type ferromagnetic alloys using the PHASE/0 code \cite{phase} with the projector augmented wave (PAW) potential\cite{1994Blochl-PRB} including the spin-orbit interaction. The spin-polarized generalized gradient approximation was adopted for the exchange and correlation energy\cite{1996Perdew-PRB}. 
 The cut-off energies of the plane-wave and the charge density were set to 70Ry and 700Ry, respectively.  The spin-orbit interaction was included by adding the spin-orbit Hamiltonian to the nonlocal PAW potential part in the DFT calculations with plane wave basis sets\cite{2016Steiner-PRB}.   We used the tetragonal unit cell for L1$_0$-$X$Pt($X$=Fe, Co and Ni) as shown in Fig. 1(c), which includes one $X$ atom and one Pt in the unit-cell. The lattice parameters of $c$ along the \mbox{\boldmath $\hat{z}$}-direction and $a$ along the \mbox{\boldmath $\hat{x}$} or \mbox{\boldmath $\hat{y}$}-direction were determined by structure optimization calculations.
We used the lattice constants $a=2.7437{\rm \AA}$ and  $c=3.7641{\rm \AA}$ for FePt, $a=2.6925{\rm \AA}$ and  $c=3.7286{\rm \AA}$ for CoPt and, $a=2.7407{\rm \AA}$ and  $c=3.5945{\rm \AA}$ for NiPt.

\begin{table}
\caption{ Number of valence electrons $N_{\rm v}$, anomalous Hall conductivity $\sigma_{\rm xy}^{\rm AHC}$, magnetization \mbox{\boldmath $\hat{m}$} dependent spin anomalous Hall conductivity (SAHC) $\sigma_{\rm xy}^{\rm SAHC+SHC}$, spin Hall conductivity (SHC) $\sigma_{\rm xy}^{\rm SHC}$, SAHC $\sigma_{\rm xy}^{\rm SAHC}$, and Hall conductivity polarization $\zeta=\frac{e}{\hbar}\sigma_{\rm xy}^{\rm SAHC}/\sigma_{\rm xy}^{\rm AHC}$, which are calculated according to Eqs(1)-(3). The units of AHC and (S)AHC are $(\Omega{\rm cm})^{-1}$  and $(\frac{\hbar}{e})(\Omega{\rm cm})^{-1}$ , respectively.}
\begin{ruledtabular}
\begin{tabular}{cccccccccc}
   &$N_{\rm v}$ & $\sigma_{\rm xy}^{\rm AHC}$ & $\sigma_{\rm xy}^{\rm SAHC+SHC}$ & $\sigma_{\rm xy}^{\rm SHC}$ & $\sigma_{\rm xy}^{\rm SAHC}$  & $\zeta$  \\
FePt& 18 & 1031 & 445 & 163 & 282  & 0.273 \\
CoPt& 19   & 481  & 563  & 115 & 448 & 0.931  \\
NiPt&  20  &  -826 & 2371 &   378  &  1993  & -2.41  \\
\end{tabular}
\end{ruledtabular}
\end{table}

To calculate the intrinsic transverse Hall conductivity,  we assume that the electric field is applied along the [100] direction (\mbox{\boldmath $\hat{x}$}), the Hall current flows in the [010] direction (\mbox{\boldmath $\hat{y}$}), and the magnetization is directed along [001] (\mbox{\boldmath $\hat{z}$}) when the anomalous Hall effect (AHE) is considered (see Figs. 1(a) and (b)).  
The anomalous Hall conductivity (AHC) and the spin Hall conductivity (SHC) can be obtained by using linear response theory for electronic conductivity \cite{1956Nakano-PTP,1957Kubo-JPSJ} as follows,
\begin{equation}
\sigma _{\rm xy}^{\alpha}=\frac{B_{\alpha}}{V}  \sum _{\scalebox{0.7}{\mbox{\boldmath $k$}}}  \Omega ^{\alpha}_{\rm xy}(\mbox{\boldmath $k$}),
\end{equation}
$\Omega ^{\alpha}_{\rm xy}(\mbox{\boldmath $k$})$ is the so-called Berry curvature of the charge or spin\cite{1982Thouless-PRL,1984Berry-PRSL}, which is given by,
\begin{equation}
\Omega^{\alpha} _{\rm xy}(\mbox{\boldmath $k$})=\frac{2\hbar ^2}{m_{\rm e}^2}\sum _{n' > n} (f_{\scalebox{0.7}{\mbox{\boldmath $k$}}n} - f_{\scalebox{0.7}{\mbox{\boldmath $k$}}n'}) \frac{{\rm Im}\langle \mbox{\boldmath $k$}n |p_{\rm x}^{\alpha}|\mbox{\boldmath $k$}n'\rangle \langle \mbox{\boldmath $k$}n'|p_{\rm y}|\mbox{\boldmath $k$}n\rangle}{(\epsilon _{\scalebox{0.7}{\mbox{\boldmath $k$}}n'}-\epsilon _{\scalebox{0.7}{\mbox{\boldmath $k$}}n})^2},
\end{equation}
where $n$ and $n'$ are the band indexes of occupied and unoccupied states. $V$, $m_{\rm e}$, $\epsilon _{\scalebox{0.7}{\mbox{\boldmath $k$}}n}$ and $f_{\scalebox{0.7}{\mbox{\boldmath $k$}}n}$ are the cell volume, electron mass, band energy and occupation function for each k-point \mbox{\boldmath $k$}, band $n$. $p_{\rm x}^{\alpha}$ and $p_{\rm y}$ are the momentum operators for the x and y directions. $\alpha$ denotes 'charge' or 'spin' current, where $p_{\rm x}^{\rm charge}=p_{\rm x}$ and  $p_{\rm x}^{\rm spin}=(p_{\rm x} s_{\rm z}+s_{\rm z} p_{\rm x}$)/2, respectively. $s_{\rm z}$ is the Pauli spin matrix.  Our definition of the momentum operator of spin current is consistent with Eq. (11) in Ref. \cite{2008Tanaka-PRB} and Eq. (23) of Ref. \cite{2019Amin-PRB}. $B_{\alpha}=e^2/\hbar$ for $\alpha$='charge' and $B_{\alpha}=e$ for $\alpha$='spin'.    $p_{\beta}=(m_e /\hbar) \partial H/\partial k_{\beta} (\beta = {\rm x,y})$, whrer $H$ is the Hamiltonian of the present system and $k_{\beta}$ is the wave-vector of the $\beta$ direction. 
$\sigma _{\rm xy}^{\rm charge}$ indicates the AHC.   $\sigma _{\rm xy}^{\rm spin}$ indicates the spin anomalous Hall conductivity (SAHC) in a ferromagnet depending on the magnetization direction \mbox{\boldmath $\hat{m}$}, i.e., $\sigma_{\rm xy}^{\rm SAHC+SHC}(\mbox{\boldmath $\hat{m}$})$, which includes \mbox{\boldmath $\hat{m}$}-dependent SAHC and \mbox{\boldmath $\hat{m}$}-independent SHC as shown in Fig. 1(a). 
 Thus, the SAHC can be obtained by,
\begin{equation}
  \sigma_{\rm xy}^{\rm SAHC}=\sigma_{\rm xy}^{\rm SAHC+SHC}\!(\mbox{\boldmath $\hat{m}$}\!\!\parallel\!\!\mbox{\boldmath $\hat{z}$})\!-\sigma_{\rm xy}^{\rm SHC}\!(\mbox{\boldmath $\hat{m}$}\!\!\parallel\!\!\mbox{\boldmath $\hat{x}$}), 
\end{equation}
 where $\mbox{\boldmath $\hat{m}$}\!\!\parallel\!\!\mbox{\boldmath $\hat{z}$}$ and $\mbox{\boldmath $\hat{m}$}\!\!\parallel\!\!\mbox{\boldmath $\hat{x}$}$ indicate perpendicular [001] and in-plane [100] magnetization directions, respectively\cite{2019Amin-PRB}.
The eigenstate $|\mbox{\boldmath $k$}n \rangle$ and the band energy $\epsilon _{\scalebox{0.7}{\mbox{\boldmath $k$}}n}$ can be obtained from the DFT calculations, including the spin-orbit interaction.  
 The convergence of  $\sigma _{\rm xy}^{\alpha}$  as a function of the k-points was carefully checked for each alloy, and 93$\times$93$\times$65 k-points in the first Brillouin zone were used for  the electronic structure calculations and linear response calculations of L1$_0$-FePt, CoPt and NiPt.

\begin{figure}[tp]

\includegraphics[height=0.18\textheight,width=0.49\textwidth]{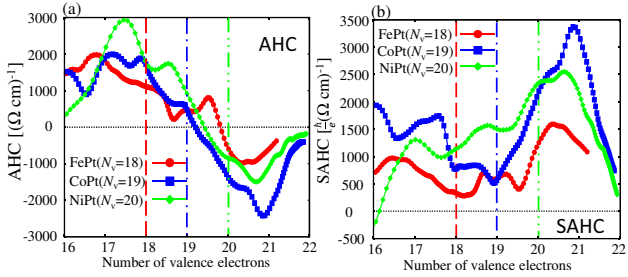}

\caption{(Color online) (a)Anomalous Hall conductivity (AHC)  $(\Omega{\rm cm})^{-1}$   and (b) Spin Anomalous Hall conductivity (SAHC)  $(\frac{\hbar}{e})(\Omega{\rm cm})^{-1}$  of L1$_0$-FePt, CoPt and NiPt as a function of the number of valence electrons $N_{\rm v}$.  The AHC and SAHC listed in Table 1 correspond to the values of $N_{\rm v}=18$ for FePt,  $N_{\rm v}=19$ for CoPt and $N_{\rm v}=20$ for NiPt, respectively.}

\end{figure}


In Table 1, we show the calculation results for  AHC $\sigma_{\rm xy}^{\rm AHC}$, SHC $\sigma_{\rm xy}^{\rm SHC}$, and SAHC $\sigma_{\rm xy}^{\rm SAHC}$ for L1$_0$-type FePt, CoPt and NiPt alloys. 
AHC and SAHC show a clear chemical trend for L1$_0$-$X$Pt ($X$=Fe,Co,Ni).
Larger AHC and smaller SAHC were obtained for FePt, while smaller AHC and larger SAHC were found for CoPt and NiPt. 
As a result, $|\zeta|$ of FePt is very small and less than 1, while the values of $|\zeta|$ for CoPt and NiPt are larger than 1. 
Thus, we can say that CoPt and NiPt are more favorable for obtaining large SOT than FePt. The small value of $\zeta$ in FePt is not consistent with the recent experimental result by Seki {\it et al.}\cite{2019Seki-PRB}, where a large $\zeta$ around 6 was estimated from the experiment. 
  
Furthermore, a longitudinal resistivity of FePt is estimated around 93$\mu \Omega{\rm cm}$ in the experiment, which corresponds to $\sigma_{\rm xx}\approx1.1\times10^4 (\Omega{\rm cm})^{-1}$. Thus, an efficiency of SAHE can be given by $\alpha_{\rm SAHE}=((\frac{e}{\hbar}\sigma_{\rm SAHC}-\beta \sigma_{\rm AHC}))⁄\sigma_{\rm xx}=((282-0.4\times1031))⁄(1.1\times10^4)\approx-0.012$. Here, we assume that a spin-polarization of the longitudinal conductivity is around 0.4, which comes from the spin-polarization of total density of states of FePt. Thus, the theoretically estimated $\alpha_{\rm SAHE}$ is one order magnitude smaller than that of the experiments ($\approx$0.25).

Since the present calculation does not include the extrinsic part of the transverse Hall conductivity, such as the skew scattering and the side jump effect due to impurities, the large value of $\zeta$ found in the experiment might be attributed to the extrinsic part of $\sigma _{\rm xy}^{\alpha}$.

To clarify the differences in AHC and SAHC for L1$_0$-$X$Pt  ($X$=Fe,Co,Ni), we show AHC and SAHC as a function of the number of valence electrons ($N_{\rm v}$) for each alloy in Fig. 2.
To obtain the $N_{\rm v}$-dependent AHC and SAHC, we evaluated the occupation function $f_{\scalebox{0.7}{\mbox{\boldmath $k$}}n}$ in equation (2) with changing $N_{\rm v}$.  
Note that AHC and SAHC shown in Table 1 correspond to the values of $N_{\rm v}=18$ for FePt,  $N_{\rm v}=19$ for CoPt and $N_{\rm v}=20$ for NiPt.
As can be seen in Fig. 2, the AHC of FePt, CoPt, and NiPt show similar valence dependence. The same is true for SAHC.
 The AHC roughly decreases with increasing $N_{\rm v}$, and the sign of AHC changes from positive to negative around $N_{\rm v}=19\sim20$,
while the SAHC increases with increasing $N_{\rm v}$ and reaches a maximum around $N_{\rm v}=20\sim21$.
These results indicate that the differences in AHC and SAHC for FePt, CoPt and NiPt can be explained by the difference in the number of valence electrons within the rigid band model.

\begin{figure}[tp]

\includegraphics[height=0.28\textheight,width=0.49\textwidth]{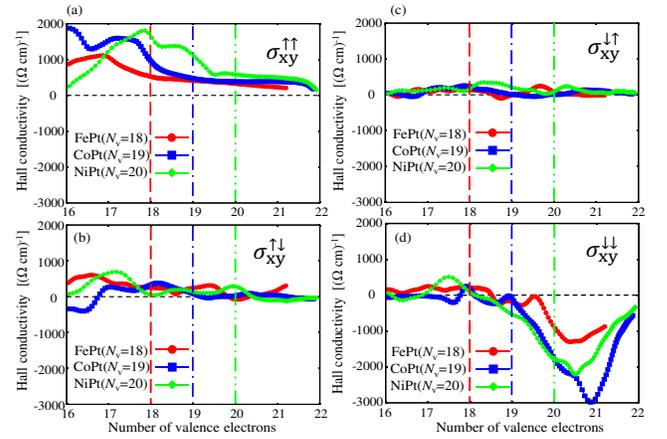}

\caption{(Color online) Spin-decomposed AHC  (a)$\sigma_{\rm xy}^{\uparrow \uparrow}$, (b)$\sigma_{\rm xy}^{\uparrow \downarrow}$, (c)$\sigma_{\rm xy}^{\downarrow \uparrow}$ and (d)$\sigma_{\rm xy}^{\downarrow \downarrow } (\Omega~{\rm cm})^{-1}$ for FePt, CoPt and NiPt as a function of the number of valence electrons $N_{\rm v}$.  }

\end{figure}

\begin{table}
\caption{ Spin-decomposed anomalous Hall conductivity $\sigma_{\rm xy}^{\uparrow \uparrow}$, $\sigma_{\rm xy}^{\downarrow \downarrow}$, $\sigma_{\rm xy}^{\uparrow \downarrow}$ and $\sigma_{\rm xy}^{\downarrow \uparrow } (\Omega~{\rm cm})^{-1}$ of FePt, CoPt and NiPt, where the anomalous Hall conductivity can be given by $\sigma_{\rm xy}^{\rm AHC}=\sigma_{\rm xy}^{\uparrow \uparrow}+\sigma_{\rm xy}^{\downarrow \downarrow}+\sigma_{\rm xy}^{\uparrow \downarrow}+\sigma_{\rm xy}^{\downarrow \uparrow }$.}
\begin{ruledtabular}
\begin{tabular}{ccccc}
$(\Omega~{\rm cm})^{-1}$   & $\sigma_{\rm xy}^{\uparrow \uparrow}$ & $\sigma_{\rm xy}^{\downarrow \downarrow}$ & $\sigma_{\rm xy}^{\uparrow \downarrow}$ & $\sigma_{\rm xy}^{\downarrow \uparrow }$   \\
FePt ($N_{\rm v}=18$)& 545 & 100 & 160 & 226    \\
CoPt ($N_{\rm v}=19$)& 472 & -91 & 85  & 15   \\
NiPt ($N_{\rm v}=20$)& 583 &  -1788 & 289 &  90    \\

\end{tabular}
\end{ruledtabular}
\end{table}

To obtain further understanding on this point, we divided the anomalous Hall conductivity for each spin component for perpendicular magnetization $\mbox{\boldmath $\hat{m}$}\!\!\parallel\!\!\mbox{\boldmath $\hat{z}$}$. Each spin component $\sigma_{\rm xy}^{\uparrow \uparrow}$, $\sigma_{\rm xy}^{\downarrow \downarrow}$, $\sigma_{\rm xy}^{\uparrow \downarrow}$ and $\sigma_{\rm xy}^{\downarrow \uparrow }$ indicates the spin component of occupied and unoccupied states in Eq. (2), which can be obtained by calculating the eigenvalues of the $s_z$ operator for each eigenstate $|\mbox{\boldmath $k$}n\rangle$ along the spin-quantum axis.  These decompositions to up-spin and down-spin states for each eigenstate are approximate because the spin-orbit interaction mixes the up-spin and down-spin states in the DFT calculations. Nonetheless, it would be very useful to understand AHC and SAHC in terms of electronic structures. 
For $\mbox{\boldmath $\hat{m}$}\!\!\parallel\!\!\mbox{\boldmath $\hat{z}$}$, the  $\sigma_{\rm xy}^{\rm AHC}$ and  $\sigma_{\rm xy}^{\rm SHC}\!(\mbox{\boldmath $\hat{m}$})$ can be given by,
\begin{equation}
\sigma_{\rm xy}^{\rm AHC}=\sigma_{\rm xy}^{\uparrow \uparrow}+\sigma_{\rm xy}^{\downarrow \downarrow}+\sigma_{\rm xy}^{\uparrow \downarrow}+\sigma_{\rm xy}^{\downarrow \uparrow },
\end{equation}
\begin{equation}
\frac{e}{\hbar}\sigma_{\rm xy}^{\rm SAHC+SHC}\!(\mbox{\boldmath $\hat{m}$}\!\!\parallel\!\!\mbox{\boldmath $\hat{z}$}) = \sigma_{\rm xy}^{\uparrow \uparrow} - \sigma_{\rm xy}^{\downarrow \downarrow}.
\end{equation}
 The Eq. (5) is consistent with the Eq. (23) in Ref. \cite{2010Naito-PRB}. It is natural that the AHC can be given by the sum of all spin components in matrix elements of momentum operator.

Table 2 shows the spin-decomposed AHC at the Fermi level for FePt, CoPt and NiPt.
In the case of FePt, all spin components have positive values, which provides larger AHC according to Eq. (4).
It was found that the spin-up-up component is dominant, and the spin-down-down component is relatively small in FePt. 
Furthermore, $\sigma_{\rm xy}^{\uparrow \downarrow}$ and $\sigma_{\rm xy}^{\downarrow \uparrow }$  show relatively large positive values, but these spin flip terms do not contribute to the spin current shown by Eq. (5).
This means that  the spin current of FePt  is dominated only by $\sigma_{\rm xy}^{\uparrow \uparrow}$ according to Eq. (5), which is much smaller than $\sigma_{\rm xy}^{\rm AHC}$.
Thus, we obtained very small Hall conductivity polarization $\zeta =\frac{e}{\hbar}\sigma_{\rm xy}^{\rm SAHC}/\sigma_{\rm xy}^{\rm AHC}$ for FePt, less than 1.
In the case of CoPt and NiPt,  the spin-up-up components are similar to that of FePt, while the spin-down-down components show negative values.
In particular, NiPt shows large negative $\sigma_{\rm xy}^{\downarrow \downarrow}$, corresponding to -1700 $(\Omega~{\rm cm})^{-1}$.
The opposite signs of $\sigma_{\rm xy}^{\uparrow \uparrow}$ and $\sigma_{\rm xy}^{\downarrow \downarrow}$ in the spin-conserving AHC cancel out the charge current in Eq. (4), but enhance the spin current in Eq. (5). 
This provides a small AHC and large SAHC, leading to a larger Hall conductivity polarization $|\zeta|$ of more than 2.4 for NiPt.
In Fig. 3, we show the spin-decomposed AHC as a function of the number of valence electrons. 
It is apparent that the spin-up-up component slightly decreases, while the spin-down-down component significantly decreases and changes sign from positive to negative with increasing number of valence electrons around $N_{\rm v}=18\sim20$.
 The change of the sign in the $\sigma_{\rm xy}^{\downarrow \downarrow}$ in Fig. 3(d) around $N_{\rm V}$=18-20 can be attributed to appearance of $d$(yz,zx) and $d$(xy) states in the minority-spin states at the Fermi level, which will be explained later. 
These trends also suggest that the behaviors of the AHC and SAHC for FePt, CoPt and NiPt could originate in the difference in the number of valence electrons within the rigid band model.

\begin{figure}[tp]

\includegraphics[height=0.5\textheight,width=0.5\textwidth]{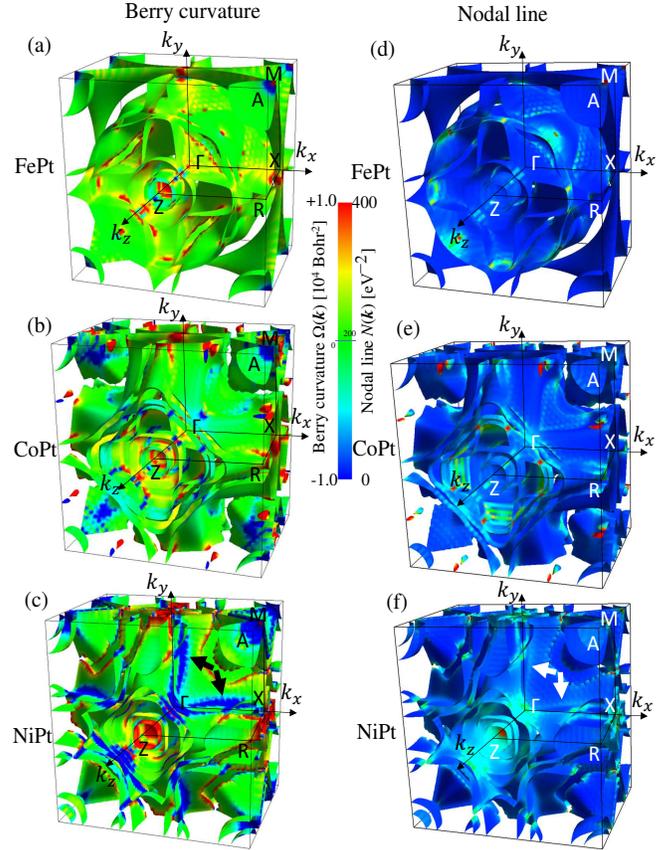}

\caption{(Color online) Color map of Berry curvatures $\Omega^{\rm charge} _{\rm xy}(\mbox{\boldmath $k$})$ in Eq. (2) for (a) FePt, (b) CoPt and (c) NiPt and the nodal line $N(\mbox{\boldmath $k$})$ in Eq. (6) for (d) FePt, (e) CoPt and (f) NiPt plotted on the Fermi surface in the three-dimensional Brillouin zone of the tetragonal unit cell, which are visualized with FermiSurfer\cite{2019Kawamura-CPC}. }

\end{figure}

Next, we consider the Berry curvature to gain insight into the large positive $\sigma_{\rm xy}^{\uparrow \uparrow}$ and negative $\sigma_{\rm xy}^{\downarrow \downarrow}$ spin, especially in CoPt and NiPt.  
In Figs. 4, we show the Berry curvature of charge $\Omega^{\rm charge} _{\rm xy}(\mbox{\boldmath $k$})$ in Eq. (2) and $N(\mbox{\boldmath $k$})$ in Eq. (6) corresponding to the denominator of Eq. (2) with occupation functions mapped onto the Fermi surface of FePt, CoPt and NiPt in the three-dimensional Brillouin zone.
\begin{equation}
N(\mbox{\boldmath $k$})=\sum _{n\neq n'} (f_{\scalebox{0.7}{\mbox{\boldmath $k$}}n} - f_{\scalebox{0.7}{\mbox{\boldmath $k$}}n'}) \frac{1}{(\epsilon _{\scalebox{0.7}{\mbox{\boldmath $k$}}n'}-\epsilon _{\scalebox{0.7}{\mbox{\boldmath $k$}}n})^2},
\end{equation}
The $N(\mbox{\boldmath $k$})$ will diverge if two eigenvalues with occupied and unoccupied states are close to each other at the same k-point.
This situation will occur when the nodal lines of band dispersions (band crossing points) are located around the Fermi level and the spin-orbit interaction opens a gap in the degenerate states. 
Thus, $N(\mbox{\boldmath $k$})$ will reflect the nodal line  of band structures without the spin-orbit interaction  in the Brillouin zone.
First, we can find that the distribution of Berry curvature $\Omega^{\rm charge} _{\rm xy}(\mbox{\boldmath $k$})$ is not the same as the distribution of the nodal line $N(\mbox{\boldmath $k$})$. 
This means that there are many k-points in which the matrix elements of the momentum operator are very small for k-points on the nodal line, due to the forbidden transition of the momentum operator.
The matrix of the momentum operator corresponding to the dipole moment operator has non-zero elements only for transitions between states of different parity. 
Thus, the value of matrix elements with the same parity is zero, where the parity corresponds to the eigenvalue of the space inversion operator for eigenstates $|\mbox{\boldmath $k$}n\rangle$.

\begin{figure}[tp]

\includegraphics[height=0.3\textheight,width=0.5\textwidth]{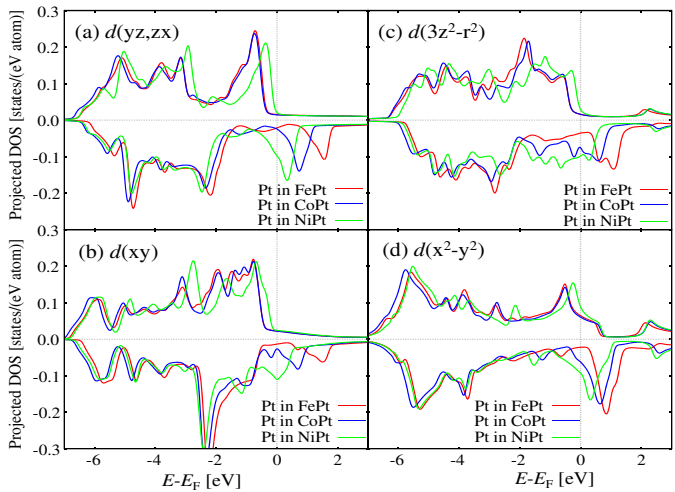}

\caption{(Color online) Projected density of states (DOS) onto each atomic orbital (a) $d({\rm yz,zx})$, (b) $d(3{\rm z}^2-{\rm r}^2)$ (c) $d({\rm xy})$ and (d) $d({\rm x}^2-{\rm y}^2)$ for FePt, CoPt and NiPt as a function of energy relative to the Fermi energy.  }

\end{figure}

In Fig. 4(a), the positive peaks of the Berry curvatures for FePt are expressed in yellow or red, especially on the $\Gamma$-Z line and the A-M line.
Positive peaks also appear along the nodal line on the Fermi surface around the  $\Gamma$-Z line  (see Figs. SM1-3 on detailed plots of $k_x-k_y, k_y-k_z$ and $k_z-k_x$ planes for the Berry curvatures and the nodal lines in the 3D Brillouin zone\cite{SM2}).
We confirmed that these positive peaks of Berry curvature come from the spin-up-up, spin-up-down and spin-down-up components, providing the large AHC of FePt.
Then, in Fig. 4(b), hot-spot-like positive and negative Berry curvatures can be found on the nodal line of CoPt.
We found that the positive peaks are from the spin-up-up component, while the negative peaks are due to the spin-down-down component in the Berry curvature of CoPt.
Since the negative spin-down-down component decreases the AHC and increases the spin current, the SAHC of CoPt exceeds the AHC, leading to the Hall polarization being larger than 1.
This trend is more remarkable in NiPt.
In Fig. 4(c), the negative Berry curvatures expressed in blue can be found on the A-M line and along the nodal line (indicated by the black and white allows in Figs. 4(c) and(f)). Some positive peaks also appear at the X point and on the $\Gamma$-Z line.
The negative Berry curvatures are due to the spin-down-down components, resulting in the huge negative SAHC of NiPt.
We also confirmed that the spin-up-down and -down-up components in the Berry curvatures of NiPt are much smaller than those of FePt and CoPt. 
 The spin-down-up component contributes to AHC, but not to the spin current.
Therefore, the SAHCs are not so large for FePt and CoPt.
These trends are consistent with the dependence of AHC and SAHC on the number of valence electrons in Figs. 2 and 3.

According to Table 2 and  Eqs. (4) and (5), a large negative $\sigma_{\rm xy}^{\downarrow \downarrow}$ is essential to obtain a large SAHC and Hall polarization $\zeta$.
The spin-orbit interaction of Pt is one order of magnitude larger than that of Fe, Co and Ni, indicating that the role of Pt atoms will be more significant than those of Fe, Co and Ni.
Thus, to clarify the origin of the negative $\sigma_{\rm xy}^{\downarrow \downarrow}$ in CoPt and NiPt from the viewpoint of electronic structures, we show in Fig. 5 the projected density of states (PDOS) on each atomic orbital of Pt atoms in FePt, CoPt and NiPt.
Since the Berry curvature and the nodal line are broadly distributed over the whole range of the Brillouin zone as shown in Fig. 4, the PDOS will be better to present the orbital properties in the Berry curvature as compared with the band dispersion along a high-symmetry line. 
We found a systematic shift in the PDOS of FePt, CoPt and NiPt and relatively large PDOS in the minority-spin states around the Fermi level for Pt $d({\rm yz,zx})$ and $d({\rm xy})$ of NiPt as compared to those of CoPt and FePt.
These minority spin states of Pt in NiPt around the Fermi level are the anti-bonding states with Ni, providing negative Berry curvatures along the nodal line on the Fermi surface, as shown in Fig. 4(c).  Note that anti-bonding properties are not directly related to the negative contribution to the Berry curvature. However, the appearance of the anti-bonding states in the minority-spin states around the Fermi level results in the negative spin down-down term for the Berry curvature of NiPt.


In summary, we have investigated the SAHC of the ferromagnetic alloys L1$_0$-FePt, CoPt and NiPt. The Hall conductivity polarization $\zeta=\frac{e}{\hbar}\sigma_{\rm xy}^{\rm SAHC}/\sigma_{\rm xy}^{\rm AHC}$ of FePt is around 0.1. This is due to the contributions of each spin component of AHC in FePt being positive. On the other hand, the SAHC increases and AHC decreases with increasing number of valence electrons ($N_{\rm v}$), leading to large Hall conductivity polarizations $|\zeta|$ of greater than 1.0 for CoPt and NiPt. Since these alloys show similar $N_{\rm v}$ dependence, the difference of FePt, CoPt and NiPt in AHC and SHC can be understood from the dependence in the number of valence electrons within the rigid band model. We found that the negative spin-down-down contribution in the AHC of CoPt and NiPt is the origin of the large SAHC and the Hall conductivity polarization $\zeta$, which is due to the anti-bonding $d({\rm yz,zx})$ and $d({\rm xy})$ orbitals of the Pt atom.

\begin{acknowledgements}
The authors are grateful to T. Seki, and T. Taniguchi for valuable discussions on this work. 
YM thanks to K. Tagami and M. Usami in asms for technical support on the PHASE/0 code.
This work was partly supported by Grant-in-Aids for Scientific Research (S) (Grant No. JP16H06332, No. JP20H00299 and No. JP20H02190) from the Ministry of Education, Culture, Sports, Science and Technology, Japan.
\end{acknowledgements}


\begin{thebibliography}{10}

\bibitem{2017Bhatti-MaterialsToday}
{S. Bhatti, R. Sbiaa, A. Hirohata, H. Ohno, S. Fukami, S. N. Piramanayagam, Materials Today {\bf 20}, 530 (2017).}

\bibitem{1989Slonczewski-PRB}
{J. C. Slonczewski, Phys. Rev. B {\bf 39}, 6995 (1989).}

\bibitem{1996Slonczewski-JMMM}
{J. C. Slonczewski, J. Mag. Mag. Mat. {\bf 159}, L1 (1996).}

\bibitem{1996Berger-PRB}
{L. Berger, Phys. Rev. B 54, 9353 (1996).}

\bibitem{1999Myers-Science}
{E. B. Myers, D. C. Ralph, J. A. Katine, R. N. Louie, R. A. Buhrman, Science 285, 6 (1999).}

\bibitem{2011Miron-Nature}
{I. M. Miron, K. Garello, G. Gaudin, P.-J. Zermatten, M. V. Costache, S. Auffret, S. Bandiera, B. Rodmacq, A. Schuhl and P. Gambardella, Nature {\bf 476}, 189-193 (2011).}

\bibitem{2012Liu-Science}
{L. Liu, C.-F. Pai, Y. Li, H. W. Tseng, D. C. Ralph, and R. A. Buhrman, Science {\bf 336}, 555 (2012).}

\bibitem{2016Fukami-NatureNano}
{S. Fukami, T. Anekawa, C. Zhang and H. Ohno, Nature Nanotech. {\bf 11}, 621 (2016).}

\bibitem{1999Hirsch-PRL}
{J. E. Hirsch, Phys. Rev. Lett. {\bf 83}, 1834 (1999).}

\bibitem{2013Hoffman-IEEETransMagn}
{A. Hoffman, IEEE Trans. Magn. {\bf 49}, 5172 (2013).}

\bibitem{2015Taniguchi-PRAp}
{T. Taniguchi, J. Grollier, and M. D. Stiles, Phys. Rev. Appl. {\bf 3}, 044001 (2015).}

\bibitem{SM1}
{See the first paragraph in Supplemental Material at [URL will be inserted by publisher] for a detailed explanation on the definition of SAHE.}

\bibitem{2017PRB-Das}
{K. S. Das, W. Y. Schoemaker, B. J. van Wees, and I. J. Vera-Marun, Phys. Rev. B {\bf 96}, 220408(R) (2017).}

\bibitem{2018Iihama-NatureElec}
{S. Iihama, T. Taniguchi, K. Yakushiji, A. Fukushima, Y. Shiota, S. Tsunegi, R. Hiramatsu, S. Yuasa, Y. Suzuki, and H. Kubota, Nat. Electron. {\bf 1}, 120 (2018).}

\bibitem{2018Gibbons-PRAp}
{J. D. Gibbons, D. MacNeill, R. A. Buhrman, and D. C. Ralph, Phys. Rev. Appl. {\bf 9}, 064033 (2018).}

\bibitem{2018Bose-PRAp}
{A. Bose, D. D. Lam, S. Bhuktare, S. Dutta, H. Singh, Y. Jibiki, M. Goto, S. Miwa, and A. A. Tulapurkar, Phys. Rev. Appl. {\bf 9}, 064026 (2018).}

\bibitem{2019Seki-PRB}
{T. Seki, S. Iihama, T. Taniguchi, and Koki Takanashi, Phys. Rev. B {\bf 100}, 144427 (2019).}

\bibitem{2019Amin-PRB}
{V. P. Amin, Junwen Li, M. D. Stiles, and P. M. Haney, Phys. Rev. B {\bf 99}, 220405(R) (2019).}

\bibitem{2021Qu-JPSJ}
{G. Qu, K. Nakamura, and M. Hayashi, J. Phys. Soc. Jpn. {\bf 90}, 024707 (2021).}
\bibitem{phase}
{Y. Yamasaki, A. Kuroda, T. Kato, J. Nara, J. Koga, T. Uda, K. Minami, and T. Ohno, Comput. Phys. Commun. {\bf 244}, 264 (2019); https://azuma.nims.go.jp/software/phase.}

\bibitem{1994Blochl-PRB}
{P. E. Bl\"ochl, Phys. Rev. B 50, 17953 (1994).}

\bibitem{1996Perdew-PRB} 
{J. P. Perdew, K. Burke, and M. Ernzerhof, Phy. Rev. Lett. {\bf 77}, 3865 (1996).}  

\bibitem{2016Steiner-PRB}
{S. Steiner, S. Khmelevskyi, M. Marsmann, and G. Kresse, Phys. Rev. B {\bf 93}, 224425 (2016).}

\bibitem{2013Miura-JPCM}
{Y. Miura, S. Ozaki, Y. Kuwahara, M. Tsujikawa, K. Abe and M. Shirai, J. Phys.: Condens. Matter 25, 106005 (2013).}

\bibitem{1956Nakano-PTP}
{ H. Nakano, Prog. Theor. Phys. {\bf 15}, 77 (1956)}
\bibitem{1957Kubo-JPSJ}
{R. Kubo, J. Phys. Soc. Jpn. {\bf 12}, 570 (1957)}

\bibitem{1982Thouless-PRL}
{D. J. Thouless, M. Kohmoto, M. P. Nightingale, and M. den Nijs, Phys. Rev. Lett. {\bf 49}, 405 (1982).}

\bibitem{1984Berry-PRSL}
{M. V. Berry, Proc. R. Soc. Lond. {\bf 392}, 45 (1984).}

\bibitem{2008Tanaka-PRB}
{T. Tanaka, H. Kontani, M. Naito, T. Naito, D. S. Hirashima, K. Yamada, and J. Inoue, Phys. Rev. B {\bf 77}, 165117 (2008).}

\bibitem{2010Naito-PRB}
{ T. Naito, D. S Hirashima, and H. Kontani, Phys. Rev. B {\bf 81}, 195111 (2010).}

\bibitem{2019Kawamura-CPC}
{M. Kawamura, Comp. Phys. Commun. {\bf 239}, 197-203 (2019).}

\bibitem{SM2}
{See Figs. SM1-3 at [URL will be inserted by publisher] for the Berry curvatures and the nodal lines in the 3D Brillouin zone.}

\end{thebibliography}
\end{document}